\documentclass[12pt,pdf,epsfig,color,png]{article}
\usepackage{amsmath}
\usepackage{amsfonts}
\usepackage{amsmath}
\usepackage{slashed}
\usepackage{amsxtra}
\usepackage{amstext}
\usepackage{amssymb}
\usepackage{color}
\usepackage{float}
\usepackage{graphicx}
\usepackage{subcaption}
\numberwithin{equation}{section}
\usepackage[dvips]{epsfig}
\usepackage{appendix}
\usepackage{youngtab}
\allowdisplaybreaks

\newcommand{\be}{\begin{equation}}
\newcommand{\ee}{\end{equation}}
\newcommand{\beq}{\begin{equation}}
\newcommand{\eeq}{\end{equation}}
\newcommand{\ba}{\begin{eqnarray}}
\newcommand{\ea}{\end{eqnarray}}
\newcommand{\bea}{\begin{eqnarray}}
\newcommand{\eea}{\end{eqnarray}}

\begin{document}
\baselineskip=15.5pt \pagestyle{plain} \setcounter{page}{1}
%
\begin{titlepage}

\vskip 0.8cm

\begin{center}

{\Large \bf Dilaton-dilatino superstring theory scattering amplitude and its Regge behavior}

\vskip 1.cm

{\large {{\bf Lucas Martin{\footnote{\tt lucasmartinar@iflp.unlp.edu.ar}}, Martin Parlanti{\footnote{\tt martin.parlanti@fisica.unlp.edu.ar}}} {\bf and  Martin
Schvellinger}{\footnote{\tt martin@fisica.unlp.edu.ar}}}}

\vskip 1.cm

{\it Instituto de F\'{\i}sica La Plata-UNLP-CONICET. \\
Boulevard 113 e 63 y 64, (1900) La Plata, Buenos Aires, Argentina \\
and \\
Departamento de F\'{\i}sica, Facultad de Ciencias Exactas,
Universidad Nacional de La Plata. \\
Calle 49 y 115, C.C. 67, (1900) La Plata, Buenos Aires, Argentina.}

\vspace{1.cm}

{\bf Abstract}

\end{center}

\vspace{.5cm}

The tree-level type II superstring theory scattering amplitude of two dilatons and two dilatinos is explicitly calculated from the Kawai-Lewellen-Tye relations. The corresponding Regge behavior is obtained.

\end{titlepage}

\newpage

{\small \tableofcontents}

\newpage

%
%
\section{Introduction}
%
%

We explicitly calculate the tree-level type II superstring theory scattering amplitude of two dilatons and two dilatinos using the Kawai-Lewellen-Tye (KLT) relations \cite{Kawai:1985xq}. These relations establish that it is possible to express an $M$-point closed string tree-level scattering amplitude as sums of products of $M$-point open string tree-level scattering amplitudes. We obtain a compact expression which holds for any scattering angle, $\theta$, as well as for the Regge limit, i.e. $\tilde{s} \gg  |\tilde{t}|$, where $\tilde{s}$, $\tilde{t}$ and $\tilde{u}$ are the Mandelstam variables.\footnote{We also show that the expression we have obtained for the tree-level type II superstring theory scattering amplitude of two dilatons and two dilatinos satisfies the ${\tilde{s}}$-${\tilde{u}}$ duality, as expected.} Besides the fact that this is a technically complicated calculation and the result is worthy on its own, for certain applications in the context of the gauge/string theory duality this expression of the scattering amplitude turns out to be essential. An example is the holographic dual calculation of the four-dimensional hard-scattering amplitude of two bosonic states created by ${\cal {O}}_{k}^{(8)}(x) \sim {\text{Tr}}(F^2_+ X_I^k)$ operators  and two fermionic states created by ${\cal {O}}_{k'}^{(6)}(x) \sim {\text{Tr}}(F_+ \lambda_{{\cal {N}}=4} X_I^{k'})$ operators of the $SU(N_c)$ ${\cal {N}}=4$ supersymmetric Yang-Mills theory developed in \cite{Martin:2025jab} using the results of the present work.\footnote{Notice that these gauge-invariant single-trace operators are built out of the gauge supermultiplet fields which are the gauge fields, the 6 real scalars $X_I$ and the gauginos $\lambda_{{\cal {N}}=4}$, transforming in the adjoint representation of the gauge group $SU(N_c)$. In addition, $k \geq 0$ and $k' \geq 0$ label the number of scalar fields.} This also may have interesting implications for  $\text{glueball}$+$\text{fermion}$ $\rightarrow$ $\text{glueball}$+$\text{fermion}$ hard scattering.

Open string theory scattering amplitudes involving two bosons and two fermions as well as four fermions were calculated long time ago \cite{Green:1981xx,Schwarz:1982jn}. Vertex operators for the emission of massless open-string states were developed \cite{Green:1981xx,Green:1981yb} using the Green-Schwarz (GS) formalism in the light-cone gauge \cite{Green:1980zg}. Four-point scattering amplitudes for massless vectors as well as spinors were calculated in  \cite{Green:1981ya,Green:1982sw}, and also one-loop scattering amplitudes were studied. In addition, open string scattering amplitudes were obtained by Atick and Sen \cite{Atick:1986rs} using covariant fermion emission vertex operators within the Ramond-Neveu-Schwarz (RNS) formalism. In reference \cite{Atick:1986rs} the corresponding closed string scattering amplitude is also presented, both at tree level and at one-loop level, as the tensor product of two open string scattering amplitudes. However, the explicit calculation of this tensor product was not carried out in \cite{Atick:1986rs}. In fact, Atick and Sen obtained an expression for the kinematic factor $K(\zeta^{(1)}, \zeta^{(2)}, u_{(3)}, u_{(4)}; k_i)$ of the closed superstring theory scattering amplitude of two bosons and two fermions  (see their equation (3.34)) in terms of the product of the kinematic factors of the scattering amplitudes of open strings. In other words, in reference \cite{Atick:1986rs} the explicit expressions of the kinematic factors of closed string scattering amplitudes were not worked out. Thus, our calculation presented in this work is the first one where, by using the KLT relations in the GS formalism\footnote{Recall that Atick and Sen used the RNS formalism  \cite{Atick:1986rs}.}, the explicit form of two bosons and two fermions scattering amplitude has been obtained in type II superstring theory. Remarkably, after a laborious work, the amplitude that we obtain has a very simple and compact form.

In addition, in reference \cite{Becker:2015eia} some tree-level string theory scattering amplitudes were studied in the RNS formalism for open strings, and then, by using the KLT relations, the authors investigated some closed string scattering amplitudes including two and four gravitinos. However, they did not included dilatinos neither obtained explicit expressions for the corresponding closed string scattering amplitudes.

Recently, we developed an explicit calculation of the tree-level four-dilaton type II superstring theory scattering amplitude \cite{Martin:2025pug}, obtaining the  Virasoro-Shapiro-Green-Schwarz-Brink amplitude \cite{Virasoro:1969me,Shapiro:1970gy,Green:1982sw}. Then, also in reference \cite{Martin:2025pug} we carried out an original and difficult calculation of the tree-level four-dilatino scattering amplitude in type II superstring theory, obtaining a long and complicated expression of this amplitude. Moreover, we used this amplitude to investigate the four-fermion hard scattering \cite{Martin:2024jpe} in the dual gauge field theory.

This work is organized as follows. In section 2 we derive the type II superstring theory scattering amplitude of two dilatons and two dilatinos by using the KLT relations. In section 3 we explicitly calculate the kinematic term which, after the tensor product of the corresponding two open string kinematic factors, leads to 270 terms that we work out explicitly. We also obtain the Regge limit of this scattering amplitude. In section 4 we present the conclusions. In appendix A we develop in detail an example of the calculation of a representative term contributing to the closed string scattering amplitude.

%
%
\section{Derivation of the dilaton-dilatino scattering amplitude from the KLT relations}
%
%

In this section we begin the derivation of the two-dilaton two-dilatino scattering amplitude using the KLT relations.
This is a type II superstring theory scattering amplitude of the form $\text{NS-NS}+\text{NS-R}\rightarrow \text{NS-NS}+\text{NS-R}${\footnote{NS-NS stands for Neveu-Schwarz-Neveu-Schwarz fields and NS-R indicates Neveu-Schwarz-Ramond fields.}}. We consider the mostly-plus ten-dimensional Minkowski space-time metric $\eta_{MN}=\text{diag}(-, +, +, \dots, +)$, where $M, N, \dots = 0, 1, \dots, 9$ are Lorentz indices, while $\alpha, \beta, \dots = 1, 2, \dots, 32$ are spinor indices.

The Clifford algebra is defined by:
\begin{eqnarray}
\{\Gamma^M,\Gamma^N\} = - 2 \eta^{MN}   \, , \label{Clifford}
\end{eqnarray}
thus, $\Gamma^M\Gamma_M=-10$. The ten-dimensional Dirac gamma matrices contain $32\times32$ purely imaginary components since they are in the Majorana representation \cite{Green:1987sp}:
\begin{eqnarray}
    \Gamma^0&=&\sigma^2 \otimes 1_{16} \, , \\
    \Gamma^j&=&i\sigma^1 \otimes \gamma^j \, , \,\,\,\,\,\,\, j=1,..,8 \\
    \Gamma^9&=&i\sigma^3 \otimes 1_{16} \, .
\end{eqnarray}
We consider the standard representation of the 16-dimensional matrices $\gamma^j$'s  given in \cite{Green:1987sp} corresponding to the reducible ${\bf {8_s}}+{\bf {8_c}}$ representation of spin(8), which can be written as:
\begin{eqnarray}
\gamma^j=\begin{pmatrix}0 & \gamma^j_{a \dot{a}} \\  \gamma^j_{\dot{b} b} & 0 
\end{pmatrix}_{16 \times 16}, 
\end{eqnarray}
being  $\gamma^j_{\dot{a} a} = (\gamma^j_{a \dot{a}})^t$.
Labels $16 \times 16$ and $8 \times 8$ indicate the dimensions of the matrices. The $\gamma^j$'s matrices must satisfy the following relations:
\begin{eqnarray}
&& \gamma^i_{a \dot{a}} \gamma^j_{\dot{a} b} + \gamma^j_{a \dot{a}} \gamma^i_{\dot{a} b} =  2 \delta^{ij} \delta_{ab} \, ,  \\
&& \gamma^i_{\dot{a} a} \gamma^j_{a \dot{b}} + \gamma^j_{\dot{a} a} \gamma^i_{a \dot{b}} =  2 \delta^{ij} \delta_{\dot{a} \dot{b}} \, .
\end{eqnarray}
Then, a complete set of eight-dimensional $\gamma^j_{a \dot{a}} \rightarrow \gamma^j_{8 \times 8}$ matrices is given by:
%
\begin{eqnarray}
&& \gamma^1_{8 \times 8} = \epsilon \times \epsilon \times \epsilon \, , \,\,\,\,\,\,\,\,\,\,\,\,\,\,\,\,\,\,\,\, \gamma^2_{8 \times 8} = I_2 \times \sigma^1 \times \epsilon \, , \nonumber \\
&& \gamma^3_{8 \times 8} = I_2 \times \sigma^3 \times \epsilon \, , \,\,\,\,\,\,\,\,\,\,\,\,\,\, \gamma^4_{8 \times 8} = \sigma^1 \times \epsilon \times I_2 \, , \nonumber \\
&& \gamma^5_{8 \times 8} = \sigma^3 \times \epsilon \times 1_2 \, , \,\,\,\,\,\,\,\,\,\,\,\,\,\, \gamma^6_{8 \times 8} = \epsilon \times I_2 \times \sigma^1 \, , \nonumber \\
&& \gamma^7_{8 \times 8} = \epsilon \times I_2 \times \sigma^3 \, , \,\,\,\,\,\,\,\,\,\,\,\,\,\, \gamma^8_{8 \times 8} = I_2 \times I_2 \times I_2 \, , 
\end{eqnarray}
where we have used that $\epsilon \equiv i \sigma^2$. The identity matrix $I_2$ and the Pauli matrices are given in the following representation:
\begin{eqnarray}
I_2=\begin{pmatrix}1 & 0 \\ 0 & 1 
\end{pmatrix}, \,\,\,\,\,
\sigma^1=\begin{pmatrix}0 & 1 \\ 1 & 0 
\end{pmatrix}, \,\,\,\,\,
\sigma^2=\begin{pmatrix}0 & -i \\ i & 0\end{pmatrix},\,\,\,\,\,
\sigma^3=\begin{pmatrix}1 & 0\\ 0& -1 
\end{pmatrix}\, .
\end{eqnarray}

The Mandelstam variables are defined as:
\begin{eqnarray}
\tilde{s}&=&-(k_1+k_2)^2 \, , \nonumber \\
\tilde{t}&=&-(k_1+k_4)^2 \, , \nonumber \\
\tilde{u}&=&-(k_1+k_3)^2 \, , 
\end{eqnarray}
assuming that all momenta $k_i$ are incoming.

From the tensor product of the open string polarizations $\zeta_i^M$ and $\zeta_i^{M'}$, which correspond to the vector polarizations of the left-moving and right-moving pieces of the closed string scattering polarization tensor
$\Theta_i^{MM'}$, it is possible to make the replacement:
\begin{eqnarray}
   \zeta_i^M\otimes \zeta_i^{M'} \rightarrow \Theta_i^{MM'} \, , \label{zeta-zeta}
\end{eqnarray}
being the transverse diagonal piece of this tensor corresponding to the dilaton polarization \cite{Gross:1986mw}:
\begin{eqnarray}
   \Theta_i^{MM'}=\frac{1}{\sqrt{8}} \left(\eta^{MM'}-k_i^M\bar{k}_i^{M'}-k_i^{M'}\bar{k}_i^{M}\right) \Phi_i \, ,
\label{theta}
\end{eqnarray} 
with $k\cdot \bar{k}=1$ and $\bar{k}\cdot\bar{k}=0$. 

In addition, the dilatino $\lambda_i^\beta$ is defined as follows \cite{Garousi:1996ad}:
\begin{eqnarray}
   u_i^\alpha \otimes \zeta_i^M = i (\Gamma^M)^\alpha_\beta \lambda_i^\beta \, ,
\end{eqnarray}
being $u_i^\alpha$ the polarization of the Majorana-Weyl spinor, with the spinor indices $\alpha$, $\beta = 1, \dots 32$. 

We calculate the following kinematic factor of the closed string scattering amplitude:
\begin{eqnarray}
    K_{\text{closed}}^{\text{2 dilatons-2 dilatinos}}(\tilde{1},2,3,\tilde{4})= \frac{1}{16}K_{\text{open}}^{\text{bosonic}}(1,2,3,4) \otimes  \frac{1}{8}K_{\text{open}}^{\text{fermionic}}(\tilde{1},2,\tilde{4},3) \label{Kc} \, ,
\end{eqnarray}
where the numerical factors correspond to the definitions:
 \begin{eqnarray}
     \frac{1}{16}K_{\text{open}}^{\text{bosonic}}(1,2,3,4)&=&K_{\text{open}}^{\text{bosonic}}(k_1/2,k_2/2,k_3/2,k_4/2) \, , \label{eq215} \\
     \frac{1}{8}K_{\text{open}}^{\text{fermionic}}(\tilde{1},2,\tilde{4},3)&=&K_{\text{open}}^{\text{fermionic}}(k_1/2,k_2/2,k_4/2,k_3/2) \, . \label{eq216}
 \end{eqnarray}
Notice that in the second factor of (\ref{Kc}) the strings 3 and 4 are exchanged in comparison with the first factor. Also, in the open-string kinematic factor one must replace the ten-dimensional momenta $k_i^M$ corresponding to each closed string by half of them, as required by the KLT relations. This is shown in equations (\ref{eq215}) and (\ref{eq216}) where we have included all numerical factors. Now, we may rewrite the kinematic factor of four open bosonic string states as \cite{Green:1987sp}:
\begin{eqnarray}
  K_{\text{open}}^{\text{bosonic}}(1,2,3,4)&=&-\frac{1}{4}(\tilde{s}\tilde{u}\zeta_1\cdot \zeta_4 \zeta_3\cdot \zeta_2+\tilde{s}\tilde{t}\zeta_1\cdot \zeta_3 \zeta_4\cdot \zeta_2+\tilde{t}\tilde{u}\zeta_1\cdot \zeta_2 \zeta_4\cdot \zeta_3) \nonumber \\
    &+&\frac{1}{2}\tilde{s}(\zeta_1\cdot k_4 \zeta_3\cdot k_2\zeta_4\cdot \zeta_2+\zeta_2\cdot k_3 \zeta_4\cdot k_1\zeta_1\cdot \zeta_3 \nonumber\\
    &+&\zeta_1\cdot k_3 \zeta_4\cdot k_2\zeta_3\cdot \zeta_2+\zeta_2\cdot k_4 \zeta_3\cdot k_1\zeta_1\cdot \zeta_4)\nonumber\\
    &+& \frac{1}{2}\tilde{t}(\zeta_2\cdot k_1 \zeta_4\cdot k_3\zeta_3\cdot \zeta_1+\zeta_3\cdot k_4 \zeta_1\cdot k_2\zeta_2\cdot \zeta_4\nonumber\\
    &+&\zeta_2\cdot k_4 \zeta_1\cdot k_3\zeta_3\cdot \zeta_4+\zeta_3\cdot k_1 \zeta_4\cdot k_2\zeta_2\cdot \zeta_1)\nonumber\\
    &+& \frac{1}{2}\tilde{u}(\zeta_1\cdot k_2 \zeta_4\cdot k_3\zeta_3\cdot \zeta_2+\zeta_3\cdot k_4 \zeta_2\cdot k_1\zeta_1\cdot \zeta_4\nonumber\\
    &+&\zeta_1\cdot k_4 \zeta_2\cdot k_3\zeta_3\cdot \zeta_4+\zeta_3\cdot k_2 \zeta_4\cdot k_1\zeta_1\cdot \zeta_2)  \, ,
\end{eqnarray}
and in the fermionic case we write:
\begin{eqnarray}
    K_{\text{open}}^{\text{fermionic}}(\tilde{1},2,\tilde{4},3)&=&\frac{\tilde{u}}{2}\bar{u}_1\Gamma \cdot \zeta_2\Gamma\cdot(k_3+k_4)\Gamma\cdot\zeta_3 u_4\nonumber\\
    &+&\frac{\tilde{s}}{2}\bar{u}_1\Gamma \cdot\zeta_3\Gamma\cdot(k_2+k_4)\Gamma\cdot\zeta_2 u_4 \, , 
\end{eqnarray}
Thus, we have to calculate:
 \begin{eqnarray}
         K_{\text{closed}}^{\text{2 dilatons-2 dilatinos}}(\tilde{1},2,3,\tilde{4})=\sum_{i=1}^{15} \left( \mathcal{T}_{i,1}+\mathcal{T}_{i,2}\right) \, ,
 \end{eqnarray}
which comes from the tensor product of the bosonic and fermionic open string kintematic factors, which in total leads to $15 \times 2=30$ terms. From each product $\zeta_i^M\otimes \zeta_i^{M'}$ there are 3 terms as indicated in the right hand side of equation (\ref{theta}) for each dilaton $\Phi_i$. Since there are 2 dilatons in the scattering amplitude, they contribute with $3 \times 3$ terms in total. Thus, multiplying $30 \times 9$ the total number of single terms is 270, all of them containing 2 dilatons, 2 dilatinos, and the corresponding ten-dimensional Dirac gamma matrices.

%
%
\section{Explicit calculation of the kinematic term, the scattering amplitude and its Regge limit}
%
%

Within the explicit calculations we use several times that 
\begin{eqnarray}
    \bar{\lambda}_1\Phi_2\Phi_3   (k_1\cdot \Gamma)\lambda_4&=&0 \, , 
    \label{dirac1} 
    \end{eqnarray}
    and
    \begin{eqnarray}
    \bar{\lambda}_1\Phi_2\Phi_3 (k_4\cdot \Gamma)\lambda_4&=&0 \, ,
    \label{dirac4}
\end{eqnarray}
since the following Dirac's equations are satisfied:
\begin{eqnarray}
    (k_4\cdot \Gamma)\lambda_4&=&0 \, ,\\
    \bar{\lambda}_1(k_1\cdot \Gamma)&=&0  \, .
\end{eqnarray}
On the other hand, we use the following conditions for the 
auxiliary momenta $\bar{k}_i$ which emerge from the dilatons polarizations:
%
\begin{eqnarray}
   \bar{\lambda}_1\Phi_2\Phi_3   \bar{k}_2\cdot \Gamma \lambda_4&=& 0 \, , \\
   \bar{\lambda}_1\Phi_2\Phi_3   \bar{k}_3\cdot \Gamma\lambda_4&=&0 \, .
\end{eqnarray}
Also, the ten-momentum conservation leads to:
\begin{eqnarray}
    k^M_1 + k^M_2 + k^M_3 + k^M_4=0 \, . \label{conservaimpulso}
\end{eqnarray}
Therefore, we have:
\begin{eqnarray}
    \bar{\lambda}_1\Phi_2\Phi_3 (k_2\cdot \Gamma)\lambda_4&=& -\bar{\lambda}_1\Phi_2\Phi_3 (k_3\cdot \Gamma)\lambda_4 \, ,
\end{eqnarray}
where we have used the momentum conservation (\ref{conservaimpulso}) together with equations (\ref{dirac1}) and (\ref{dirac4}). This result is important since it allows to express the full kinematic factor of the closed string scattering amplitude just in terms of 
$k_3$, which is the ten-dimensional momentum associated with the dilaton $\Phi_3$, which leads to simplifications during the explicit calculations.

There are terms involving the product of three Dirac gamma matrices which, by using the momentum conservation (\ref{conservaimpulso}), can be reduced to the form:
\begin{eqnarray}
    \tilde{s}^2(k_2\cdot \Gamma)(k_2 \cdot \Gamma)(k_3 \cdot \Gamma) \, .
\end{eqnarray}
We can show that this term vanishes by using the  
Clifford algebra (\ref{Clifford}) and the fact that 
dilatons and dilatinos are massless, i.e. $k_i \cdot k_{i}=0$:
\begin{eqnarray}
    \tilde{s}^2(k_2\cdot \Gamma)(k_2\cdot \Gamma)(k_3\cdot \Gamma) &=&  \tilde{s}^2 k_{2M}  k_{2N}k_{3P} \Gamma^M\Gamma^N  \Gamma^P \nonumber\\
    &=& -\tilde{s}^2 k_{2M} k_{2N}k_{3P}\left[ -2\eta^{MN} \Gamma^P+ \Gamma^N\Gamma^M  \Gamma^P \right] \nonumber\\
    &=&- \tilde{s}^2 \left[-2k_2\cdot k_2k_3\cdot \Gamma + (k_2\cdot \Gamma)(k_2\cdot \Gamma)(k_3\cdot \Gamma) \right] \nonumber\\ 
    &=& - \tilde{s}^2(k_2\cdot \Gamma)(k_2\cdot \Gamma)(k_3\cdot \Gamma) \label{k2k2k3}
\end{eqnarray}
which implies that terms containing the product of three Dirac gamma matrices of the form 
$(k_i\cdot \Gamma)(k_j\cdot \Gamma)(k_j\cdot \Gamma)$
vanish. Also, the Clifford algebra 
(\ref{Clifford}) allows to recast many other terms with three Dirac gamma matrices as terms with only one Dirac gamma matrix. For instance we have:
\begin{eqnarray}
\bar{\lambda}_1\Phi_2\Phi_3   (k_2\cdot \Gamma) (k_3\cdot \Gamma)(k_2\cdot \Gamma)\lambda_4 &=& -2 k_2 \cdot k_3  \bar{\lambda}_1\Phi_2\Phi_3   (k_2\cdot \Gamma) \lambda_4 \nonumber \\
&& - \bar{\lambda}_1\Phi_2\Phi_3   (k_3\cdot \Gamma) (k_2\cdot \Gamma)(k_2\cdot \Gamma)\lambda_4 \nonumber\\
&=&-2 k_2 \cdot k_3  \bar{\lambda}_1\Phi_2\Phi_3   (k_2\cdot \Gamma) \lambda_4 \nonumber \\
&=& \nonumber \tilde{t} \bar{\lambda}_1\Phi_2\Phi_3   (k_2\cdot \Gamma) \lambda_4 \nonumber \\
&=& -\tilde{t} \bar{\lambda}_1\Phi_2\Phi_3   (k_3\cdot \Gamma) \lambda_4 \, .
\end{eqnarray}
This notably simplifies the calculations as shown in the following example: 
\begin{eqnarray}
 &&   \bar{\lambda}_1\Phi_2\Phi_3   (k_4\cdot \Gamma) (k_3\cdot \Gamma) (k_2-k_3)\cdot \Gamma\lambda_4 \nonumber \\
    &=& \bar{\lambda}_1\Phi_2\Phi_3   (-k_1-k_2-k_3)\cdot \Gamma (k_3\cdot \Gamma) (k_2-k_3)\cdot \Gamma\lambda_4 \nonumber\\
    &=& -\bar{\lambda}_1\Phi_2\Phi_3   (k_2+k_3)\cdot \Gamma (k_3\cdot \Gamma) (k_2-k_3)\cdot \Gamma\lambda_4 \nonumber \\
    &=& \bar{\lambda}_1\Phi_2\Phi_3   (k_2+k_3)\cdot \Gamma (k_3\cdot \Gamma) (k_3-k_2)\cdot \Gamma\lambda_4 \nonumber\\
     &=& \bar{\lambda}_1\Phi_2\Phi_3   (k_2\cdot \Gamma) (k_3\cdot \Gamma)(k_3\cdot \Gamma)\lambda_4 - \bar{\lambda}_1\Phi_2\Phi_3   (k_2\cdot \Gamma) (k_3\cdot \Gamma)(k_2\cdot \Gamma)\lambda_4 \nonumber \\
     &~& + \bar{\lambda}_1\Phi_2\Phi_3   (k_3\cdot \Gamma) (k_3\cdot \Gamma)(k_3\cdot \Gamma)\lambda_4 - \bar{\lambda}_
     1\Phi_2\Phi_3   (k_3\cdot \Gamma) (k_3\cdot \Gamma)(k_2\cdot \Gamma)\lambda_4 \nonumber \\
     &=& - \bar{\lambda}_1\Phi_2\Phi_3   (k_2\cdot \Gamma) (k_3\cdot \Gamma)(k_2\cdot \Gamma)\lambda_4 \nonumber \\
     &=& 2 k_2 \cdot k_3  \bar{\lambda}_1\Phi_2\Phi_3   (k_2\cdot \Gamma) \lambda_4 + \bar{\lambda}_1\Phi_2\Phi_3   (k_3\cdot \Gamma) (k_2\cdot \Gamma)(k_2\cdot \Gamma)\lambda_4 \nonumber \\
     &=& -\tilde{t}  \bar{\lambda}_1\Phi_2\Phi_3   (k_2\cdot \Gamma) \lambda_4 = \tilde{t}  \bar{\lambda}_1\Phi_2\Phi_3   (k_3\cdot \Gamma) \lambda_4 \, . 
\end{eqnarray}

Next we introduce the final form of each of the 30 terms of $K_{\text{closed}}^{\text{2 dilatons-2 dilatinos}}(\tilde{1},2,3,\tilde{4})$. The terms of the form $\mathcal{T}_{1, j}$ are given by:
\begin{eqnarray}
  \mathcal{T}_{1,1}&=&   -\frac{1}{4\cdot 16}(\tilde{s}\tilde{u}\zeta_1\cdot \zeta_4 \zeta_3\cdot \zeta_2)\otimes \frac{\tilde{u}}{2\cdot 8}\bar{u}_1\Gamma \cdot \zeta_2 \Gamma \cdot(k_3+k_4)\Gamma\cdot \zeta_3u_4\\
     &=& -\frac{\tilde{s}\tilde{u}^2}{128}\bar{\lambda}_1\Phi_2\Phi_3 (k_3\cdot\Gamma)  \lambda_4,\nonumber\\
 \mathcal{T}_{1,2}&=&  -\frac{1}{4\cdot 16}(\tilde{s}\tilde{u}\zeta_1\cdot \zeta_4 \zeta_3\cdot \zeta_2)\otimes\frac{\tilde{s}}{2\cdot 8}\bar{u}_1\Gamma\cdot\zeta_3\Gamma\cdot (k_2+k_4)\Gamma\cdot\zeta_2u_4\\
&=& \frac{\tilde{s}^2\tilde{u}}{128}\bar{\lambda}_1\Phi_2\Phi_3 (k_3\cdot\Gamma)  \lambda_4 \nonumber.
\end{eqnarray}
The terms of the form $\mathcal{T}_{2, j}$ are:
\begin{eqnarray}
\mathcal{T}_{2,1}&=& -\frac{1}{4\cdot 16}(\tilde{s}\tilde{t}\zeta_1\cdot \zeta_3 \zeta_4\cdot \zeta_2)\otimes\frac{\tilde{u}}{2\cdot 8}\bar{u}_1\Gamma \cdot \zeta_2 \Gamma \cdot(k_3+k_4)\Gamma\cdot \zeta_3u_4\\
&=&  \frac{11\tilde{s}\tilde{t}\tilde{u}}{2048}\bar{\lambda}_1\Phi_2\Phi_3 (k_3\cdot\Gamma) \lambda_4, \nonumber\\
\mathcal{T}_{2,2}&=& -\frac{1}{4\cdot 16}(\tilde{s}\tilde{t}\zeta_1\cdot \zeta_3 \zeta_4\cdot \zeta_2)\otimes\frac{\tilde{s}}{2\cdot 8}\bar{u}_1\Gamma\cdot\zeta_3\Gamma\cdot (k_2+k_4)\Gamma\cdot\zeta_2u_4 \\
&=&  \frac{25\tilde{s^2}\tilde{t}}{2048} \bar{\lambda}_1\Phi_2\Phi_3 (k_3\cdot\Gamma)  \lambda_4\nonumber.
\end{eqnarray}
The terms of the form $\mathcal{T}_{3, j}$ are:
\begin{eqnarray}
 \mathcal{T}_{3,1}&=&  -\frac{1}{4\cdot 16}(\tilde{t}\tilde{u}\zeta_1\cdot \zeta_2 \zeta_4\cdot \zeta_3)\otimes\frac{\tilde{u}}{2\cdot 8}\bar{u}_1\Gamma \cdot \zeta_2 \Gamma \cdot(k_3+k_4)\Gamma\cdot \zeta_3u_4\\
&=&  -\frac{25\tilde{t}\tilde{u}^2}{2048}\bar{\lambda}_1\Phi_2\Phi_3 (k_3\cdot\Gamma) \lambda_4,\nonumber\\ 
\mathcal{T}_{3,2}&=&      -\frac{1}{4\cdot 16}(\tilde{t}\tilde{u}\zeta_1\cdot \zeta_2 \zeta_4\cdot \zeta_3)\otimes\frac{\tilde{s}}{2\cdot 8}\bar{u}_1\Gamma\cdot\zeta_3\Gamma\cdot (k_2+k_4)\Gamma\cdot\zeta_2u_4\\
&=& -\frac{11\tilde{s}\tilde{t}\tilde{u}}{2048} \bar{\lambda}_1\Phi_2\Phi_3 (k_3\cdot\Gamma)  \lambda_4\nonumber.
\end{eqnarray}
The terms of the form $\mathcal{T}_{4, j}$ are:
\begin{eqnarray}
\mathcal{T}_{4,1}&=&     \frac{\tilde{s}}{2\cdot 16}(\zeta_1\cdot k_4 \zeta_3\cdot k_2\zeta_4\cdot \zeta_2)\otimes\frac{\tilde{u}}{2\cdot 8}\bar{u}_1\Gamma \cdot \zeta_2 \Gamma \cdot(k_3+k_4)\Gamma\cdot \zeta_3u_4\\
&=&  -\frac{3\tilde{s}\tilde{t}\tilde{u}}{2048} \bar{\lambda}_1\Phi_2\Phi_3(k_3\cdot\Gamma) \lambda_4,\nonumber\\
\mathcal{T}_{4,2}&=&      \frac{\tilde{s}}{2\cdot 16}(\zeta_1\cdot k_4 \zeta_3\cdot k_2\zeta_4\cdot \zeta_2)\otimes\frac{\tilde{s}}{2\cdot 8}\bar{u}_1\Gamma\cdot\zeta_3\Gamma\cdot (k_2+k_4)\Gamma\cdot\zeta_2u_4\\
&=&\frac{5\tilde{s}^2\tilde{t}}{2048} \bar{\lambda}_1\Phi_2\Phi_3(k_3\cdot\Gamma) \lambda_4\nonumber. 
\end{eqnarray}
The terms of the form $\mathcal{T}_{5, j}$ are:
\begin{eqnarray}
\mathcal{T}_{5,1} &=&\frac{\tilde{s}}{2\cdot 16}(\zeta_2\cdot k_3 \zeta_4\cdot k_1\zeta_1\cdot \zeta_3)\otimes\frac{\tilde{u}}{2\cdot 8}\bar{u}_1\Gamma \cdot \zeta_2 \Gamma \cdot(k_3+k_4)\Gamma\cdot \zeta_3u_4\\
&=& -\frac{3\tilde{s}\tilde{t}\tilde{u}}{2048} \bar{\lambda}_1\Phi_2\Phi_3(k_3\cdot\Gamma) \lambda_4, \nonumber \\
 \mathcal{T}_{5,2}  &=&    \frac{\tilde{s}}{2\cdot 16}(\zeta_2\cdot k_3 \zeta_4\cdot k_1\zeta_1\cdot \zeta_3)\otimes\frac{\tilde{s}}{2\cdot 8}\bar{u}_1\Gamma\cdot\zeta_3\Gamma\cdot (k_2+k_4)\Gamma\cdot\zeta_2u_4,\\
 &=&\frac{5\tilde{s}^2\tilde{t}}{2048} \bar{\lambda}_1\Phi_2\Phi_3(k_3\cdot\Gamma) \lambda_4\nonumber.
\end{eqnarray}
The terms of the form $\mathcal{T}_{6, j}$ are:
\begin{eqnarray}
\mathcal{T}_{6,1}&=&     \frac{\tilde{s}}{2\cdot 16}(\zeta_1\cdot k_3 \zeta_4\cdot k_2\zeta_3\cdot \zeta_2)\otimes\frac{\tilde{u}}{2\cdot 8}\bar{u}_1\Gamma \cdot \zeta_2 \Gamma \cdot(k_3+k_4)\Gamma\cdot \zeta_3u_4\\
 &=& \frac{\tilde{s}\tilde{u}^2}{512} \bar{\lambda}_1\Phi_2\Phi_3 (k_3\cdot\Gamma) \lambda_4,\nonumber\\
\mathcal{T}_{6,2}&=&\frac{\tilde{s}}{2\cdot 16}(\zeta_1\cdot k_3 \zeta_4\cdot k_2\zeta_3\cdot \zeta_2)\otimes \frac{\tilde{s}}{2\cdot 8}\bar{u}_1\Gamma\cdot\zeta_3\Gamma\cdot (k_2+k_4)\Gamma\cdot\zeta_2u_4\\
&=&\frac{\tilde{s}^2\tilde{u}}{512} \bar{\lambda}_1\Phi_2\Phi_3 (k_3\cdot\Gamma) \lambda_4\nonumber.
\end{eqnarray}
The terms of the form $\mathcal{T}_{7, j}$ are:
\begin{eqnarray}
  \mathcal{T}_{7,1}&=&\frac{\tilde{s}}{2\cdot 16}(\zeta_2\cdot k_4 \zeta_3\cdot k_1\zeta_1\cdot \zeta_4)\otimes\frac{\tilde{u}}{2\cdot 8}\bar{u}_1\Gamma \cdot \zeta_2 \Gamma \cdot(k_3+k_4)\Gamma\cdot \zeta_3u_4\\
    &=&  \frac{\tilde{s}\tilde{u}^2}{512} \bar{\lambda}_1\Phi_2\Phi_3 (k_3\cdot\Gamma) \lambda_4,\nonumber\\
 \mathcal{T}_{7,2}&=&\frac{\tilde{s}}{2\cdot 16}(\zeta_2\cdot k_4 \zeta_3\cdot k_1\zeta_1\cdot \zeta_4)\otimes \frac{\tilde{s}}{2\cdot 8}\bar{u}_1\Gamma\cdot\zeta_3\Gamma\cdot (k_2+k_4)\Gamma\cdot\zeta_2u_4 \\
     &=& -\frac{\tilde{s}^3}{512} \bar{\lambda}_1\Phi_2\Phi_3 (k_3\cdot\Gamma) \lambda_4\nonumber.
\end{eqnarray}
The terms of the form $\mathcal{T}_{8, j}$ are:
\begin{eqnarray}
  \mathcal{T}_{8,1}&=&    \frac{\tilde{t}}{2\cdot 16}(\zeta_2\cdot k_1 \zeta_4\cdot k_3\zeta_3\cdot \zeta_1)\otimes\frac{\tilde{u}}{2\cdot 8}\bar{u}_1\Gamma \cdot \zeta_2 \Gamma \cdot(k_3+k_4)\Gamma\cdot \zeta_3u_4\\
     &=& -\frac{3\tilde{s}\tilde{t}\tilde{u}}{2048} \bar{\lambda}_1\Phi_2\Phi_3  (k_3\cdot\Gamma) \lambda_4 \nonumber,\\
  \mathcal{T}_{8,2}&=&       \frac{\tilde{t}}{2\cdot 16}(\zeta_2\cdot k_1 \zeta_4\cdot k_3\zeta_3\cdot \zeta_1)\otimes \frac{\tilde{s}}{2\cdot 8}\bar{u}_1\Gamma\cdot\zeta_3\Gamma\cdot (k_2+k_4)\Gamma\cdot\zeta_2u_4 \\
        &=& -\frac{5\tilde{s}\tilde{t}(\tilde{t}-\tilde{u})}{2048}\bar{\lambda}_1\Phi_2\Phi_3 (k_3\cdot\Gamma)  \lambda_4\nonumber.
\end{eqnarray}
The terms of the form $\mathcal{T}_{9, j}$ are:
\begin{eqnarray}
\mathcal{T}_{9,1}&=&\frac{\tilde{t}}{2\cdot 16}(\zeta_3\cdot k_4 \zeta_1\cdot k_2\zeta_2\cdot \zeta_4)\otimes\frac{\tilde{u}}{2\cdot 8}\bar{u}_1\Gamma \cdot \zeta_2 \Gamma \cdot(k_3+k_4)\Gamma\cdot \zeta_3u_4\\
     &=&  -\frac{3\tilde{s}\tilde{t}\tilde{u}}{2048} \bar{\lambda}_1\Phi_2\Phi_3 ( k_3\cdot\Gamma) \lambda_4, \nonumber\\
 \mathcal{T}_{9,2}&=& \frac{\tilde{t}}{2\cdot 16}(\zeta_3\cdot k_4 \zeta_1\cdot k_2\zeta_2\cdot \zeta_4)\otimes \frac{\tilde{s}}{2\cdot 8}\bar{u}_1\Gamma\cdot\zeta_3\Gamma\cdot (k_2+k_4)\Gamma\cdot\zeta_2u_4 \\
     &=& -\frac{5\tilde{s}\tilde{t}(\tilde{t}-\tilde{u})}{2048} \bar{\lambda}_1\Phi_2\Phi_3 (k_3\cdot\Gamma) \lambda_4.\nonumber
\end{eqnarray}
The terms of the form $\mathcal{T}_{10, j}$ are:
\begin{eqnarray}
 \mathcal{T}_{10,1}&=&\frac{\tilde{t}}{2\cdot 16}(\zeta_2\cdot k_4 \zeta_1\cdot k_3\zeta_3\cdot \zeta_4)\otimes\frac{\tilde{u}}{2\cdot 8}\bar{u}_1\Gamma \cdot \zeta_2 \Gamma \cdot(k_3+k_4)\Gamma\cdot \zeta_3u_4\\
     &=&  -\frac{5\tilde{t}\tilde{u}(\tilde{s}-\tilde{t})}{2048} \bar{\lambda}_1\Phi_2\Phi_3 (k_3\cdot\Gamma) \lambda_4,\nonumber\\
  \mathcal{T}_{10,2}&=&\frac{\tilde{t}}{2\cdot 16}(\zeta_2\cdot k_4 \zeta_1\cdot k_3\zeta_3\cdot \zeta_4)\otimes \frac{\tilde{s}}{2\cdot 8}\bar{u}_1\Gamma\cdot\zeta_3\Gamma\cdot (k_2+k_4)\Gamma\cdot\zeta_2u_4 \\
    &=& \frac{3\tilde{s}\tilde{t}\tilde{u}}{2048}  \bar{\lambda}_1\Phi_2\Phi_3 (k_3\cdot\Gamma)\lambda_4\nonumber.
\end{eqnarray}
The terms of the form $\mathcal{T}_{11, j}$ are:
\begin{eqnarray}
  \mathcal{T}_{11,1}&=& \frac{\tilde{t}}{2\cdot 16}(\zeta_3\cdot k_1 \zeta_4\cdot k_2\zeta_2\cdot \zeta_1)\otimes\frac{\tilde{u}}{2\cdot 8}\bar{u}_1\Gamma \cdot \zeta_2 \Gamma \cdot(k_3+k_4)\Gamma\cdot \zeta_3u_4\\
     &=&  -\frac{5\tilde{t}\tilde{u}(\tilde{s}-\tilde{t})}{2048} \bar{\lambda}_1\Phi_2\Phi_3 (k_3\cdot\Gamma) \lambda_4,\nonumber\\
 \mathcal{T}_{11,2}&=& \frac{\tilde{t}}{2\cdot 16}(\zeta_3\cdot k_1 \zeta_4\cdot k_2\zeta_2\cdot \zeta_1)\otimes\frac{\tilde{s}}{2\cdot 8}\bar{u}_1\Gamma\cdot\zeta_3\Gamma\cdot (k_2+k_4)\Gamma\cdot\zeta_2u_4 \\
     &=& \frac{3\tilde{s}\tilde{t}\tilde{u}}{2048} \bar{\lambda}_1\Phi_2\Phi_3 (k_3\cdot\Gamma)\lambda_4.\nonumber
\end{eqnarray}
The terms of the form $\mathcal{T}_{12, j}$ are:
\begin{eqnarray}
\mathcal{T}_{12,1}&=&     \frac{\tilde{u}}{2\cdot 16}(\zeta_1\cdot k_2 \zeta_4\cdot k_3\zeta_3\cdot \zeta_2)\otimes\frac{\tilde{u}}{2\cdot 8}\bar{u}_1\Gamma \cdot \zeta_2 \Gamma \cdot(k_3+k_4)\Gamma\cdot \zeta_3u_4\\
&=&  -\frac{\tilde{s}\tilde{u}^2}{512} \bar{\lambda}_1\Phi_2\Phi_3 (k_3\cdot\Gamma) \lambda_4, \nonumber\\
\mathcal{T}_{12,2}&=&    \frac{\tilde{u}}{2\cdot 16}(\zeta_1\cdot k_2 \zeta_4\cdot k_3\zeta_3\cdot \zeta_2)\otimes\frac{\tilde{s}}{2\cdot 8}\bar{u}_1\Gamma\cdot\zeta_3\Gamma\cdot (k_2+k_4)\Gamma\cdot\zeta_2u_4 \\
&=& -\frac{\tilde{s}^2\tilde{u}}{512} \bar{\lambda}_1\Phi_2\Phi_3 (k_3\cdot\Gamma) \lambda_4\nonumber.
\end{eqnarray}
The terms of the form $\mathcal{T}_{13, j}$ are:
\begin{eqnarray}
\mathcal{T}_{13,1}&=&      \frac{\tilde{u}}{2\cdot 16}(\zeta_3\cdot k_4 \zeta_2\cdot k_1\zeta_1\cdot \zeta_4) \otimes\frac{\tilde{u}}{2\cdot 8}\bar{u}_1\Gamma \cdot \zeta_2 \Gamma \cdot(k_3+k_4)\Gamma\cdot \zeta_3u_4\\
     &=& \frac{\tilde{u}^3}{512} \bar{\lambda}_1\Phi_2\Phi_3 (k_3\cdot\Gamma) \lambda_4,\nonumber\\
\mathcal{T}_{13,2}&=&\frac{\tilde{u}}{2\cdot 16}(\zeta_3\cdot k_4 \zeta_2\cdot k_1\zeta_1\cdot \zeta_4) \otimes\frac{\tilde{s}}{2\cdot 8}\bar{u}_1\Gamma\cdot\zeta_3\Gamma\cdot (k_2+k_4)\Gamma\cdot\zeta_2u_4 \\
&=& -\frac{\tilde{s}^2\tilde{u}}{512} \bar{\lambda}_1\Phi_2\Phi_3(k_3\cdot\Gamma)\lambda_4\nonumber.
\end{eqnarray}
The terms of the form $\mathcal{T}_{14, j}$ are:
\begin{eqnarray}
\mathcal{T}_{14,1}&=&     \frac{\tilde{u}}{2\cdot 16}(\zeta_1\cdot k_4 \zeta_2\cdot k_3\zeta_3\cdot \zeta_4) \otimes\frac{\tilde{u}}{2\cdot 8}\bar{u}_1\Gamma \cdot \zeta_2 \Gamma \cdot(k_3+k_4)\Gamma\cdot \zeta_3u_4\\
     &=&  -\frac{5\tilde{t}\tilde{u}^2}{2048} \bar{\lambda}_1\Phi_2\Phi_3 (k_3\cdot\Gamma)  \lambda_4, \nonumber\\
\mathcal{T}_{14,2}&=&   \frac{\tilde{u}}{2\cdot 16}(\zeta_1\cdot k_4 \zeta_2\cdot k_3\zeta_3\cdot \zeta_4) \otimes\frac{\tilde{s}}{2\cdot 8}\bar{u}_1\Gamma\cdot\zeta_3\Gamma\cdot (k_2+k_4)\Gamma\cdot\zeta_2u_4 \\
&=&\frac{3\tilde{s}\tilde{t}\tilde{u}}{2048} \bar{\lambda}_1\Phi_2\Phi_3 (k_3\cdot\Gamma)\lambda_4\nonumber.
\end{eqnarray}
Finally, the terms of the form $\mathcal{T}_{15, j}$ are:
\begin{eqnarray}
\mathcal{T}_{15,1}&=&   \frac{\tilde{u}}{2\cdot 16}(\zeta_3\cdot k_2 \zeta_4\cdot k_1\zeta_1\cdot \zeta_2)\otimes\frac{\tilde{u}}{2\cdot 8}\bar{u}_1\Gamma \cdot \zeta_2 \Gamma \cdot(k_3+k_4)\Gamma\cdot \zeta_3u_4\\
     &=&  -\frac{5\tilde{t}\tilde{u}^2}{2048} \bar{\lambda}_1\Phi_2\Phi_3 (k_3\cdot\Gamma) \lambda_4, \nonumber\\
\mathcal{T}_{15,2}&=&  \frac{\tilde{u}}{2\cdot 16}(\zeta_3\cdot k_2 \zeta_4\cdot k_1\zeta_1\cdot \zeta_2) \otimes\frac{\tilde{s}}{2\cdot 8}\bar{u}_1\Gamma\cdot\zeta_3\Gamma\cdot (k_2+k_4)\Gamma\cdot\zeta_2u_4 \\
&=&\frac{3\tilde{s}\tilde{t}\tilde{u}}{2048} \bar{\lambda}_1\Phi_2\Phi_3 (k_3\cdot\Gamma) \lambda_4\nonumber.
\end{eqnarray}
Thus, after carrying out the sum of all these terms we obtain:
\begin{eqnarray}
&& K_{\text{closed}}^{\text{2 dilatons-2 dilatinos}}(\tilde{1},2,3,\tilde{4}) \nonumber \\ 
&& = -\frac{\left(\tilde{s}-\tilde{u}\right)\left(4\tilde{s}^2-35\tilde{s}\tilde{t}+10\tilde{t}^2-8\tilde{s}\tilde{u}-35\tilde{t}\tilde{u}+4\tilde{u}^2\right)}{2048} \times \nonumber \\
 &&  \,\,\,\,\,\,\,\,\,\, \bar{\lambda}_1\Phi_2\Phi_3(k_3\cdot \Gamma)\lambda_4 \, , \label{eq343}
\end{eqnarray}
and using that $\tilde{t}=-\tilde{s}-\tilde{u}$ equation (\ref{eq343}) becomes:
\begin{eqnarray}
 K_{\text{closed}}^{\text{2 dilatons-2 dilatinos}}(\tilde{1},2,3,\tilde{4}) &=& \frac{(\tilde{s}-\tilde{u})}{4096}(49\tilde{s}^2+82\tilde{s}\tilde{u}+49\tilde{u}^2)\bar{\lambda}_1\Phi_2\Phi_3(k_2-k_3)\cdot \Gamma\lambda_4 \, . \nonumber \\
 &&
\end{eqnarray}
therefore $ K_{\text{closed}}^{\text{2 dilatons-2 dilatinos}}(\tilde{1},2,3,\tilde{4})$ is $\tilde{s}$-$\tilde{u}$ dual explicitly (recall that this duality implies the $\tilde{s} \leftrightarrow \tilde{u}$ exchange, which implies that $k_2 \leftrightarrow k_3$). Now, using the kinematic relations with the scattering angle:
\begin{eqnarray}
    \tilde{t}&=&-\frac{\tilde{s}}{2}(1-\cos \theta) \, ,\\
    \tilde{u}&=&-\frac{\tilde{s}}{2}(1+\cos \theta) \, ,
\end{eqnarray}
we can express this equation in terms of $\tilde{s}$ and the scattering angle $\theta$, obtaining\footnote{We note that equation (\ref{Ktheta}) contains the term $(7+ \cos(2\theta))$, which also appears in the kinematic factor of the dilaton-dilaton scattering amplitude \cite{Green:1982sw}. }:
\begin{eqnarray}
   K_{\text{closed}}^{\text{2 dilatons-2 dilatinos}}(\tilde{1},2,3,\tilde{4}) &=& -\frac{\tilde{s}^3(-321\cos(\theta)+162(7+ \cos(2\theta))+49\cos(3\theta))}{65536}   \nonumber \\
   && \times \ \bar{\lambda}_1\Phi_2\Phi_3(k_3\cdot \Gamma)\lambda_4 \label{Ktheta} \, .
\end{eqnarray}
Thus, the scattering amplitude is:
\begin{eqnarray} 
\mathcal{A}_{\text{closed}}^{\text{2 dilatons-2 dilatinos}}(\tilde{1},2,3,\tilde{4})=-\pi \kappa^2 {\alpha'}^3 K_{\text{closed}}^{\text{2 dilatons-2 dilatinos}}(\tilde{1},2,3,\tilde{4})\prod_{\chi=\tilde{s},\tilde{t},\tilde{u}}\frac{\Gamma(-\alpha'\chi/4)}{\Gamma(1-\alpha'\chi/4)} \, , \nonumber \\
&& \label{amplitude}
\end{eqnarray}
where  $\kappa=2 g_s \alpha'^2$.

Considering the kinematic situation where $\tilde{s} \gg |\tilde{t}|$ we obtain the Regge limit of the scattering amplitude (\ref{amplitude}):
\begin{eqnarray}
\mathcal{A}_{\text{closed-Regge}}^{\text{2 dilatons-2 dilatinos}}(\tilde{1},2,3,\tilde{4})&=& \pi \kappa^2 {\alpha'}^3 K_{\text{closed-Regge}}^{\text{2 dilatons-2 dilatinos}}(\tilde{1},2,3,\tilde{4}) \frac{\sin[\frac{\pi\alpha'}{4}(\tilde{s}+\tilde{t})]}{\sin(\frac{\pi\alpha'\tilde{s}}{4})}  \nonumber \\
&~& \times \ \nonumber \left(\frac{\alpha'\tilde{s}}{4}\right)^{-2+\frac{\alpha'\tilde{t}}{2}}  \ e^{2-\frac{\alpha'\tilde{t}}{2}}\frac{\Gamma(-\alpha'\tilde{t}/4)} {\Gamma(1+\alpha'\tilde{t}/4)}\nonumber \\ 
& = &-\frac{\pi g_s^2\alpha'^4}{16}\frac{\sin[\frac{\pi\alpha'}{4}(\tilde{s}+\tilde{t})]}{\sin(\frac{\pi\alpha'\tilde{s}}{4})} \left(\alpha'\tilde{s}\right)^{1+\frac{\alpha'\tilde{t}}{2}} \nonumber \\
&~& \times \ e^{2-\frac{\alpha'\tilde{t}}{2}}\frac{\Gamma(-\alpha'\tilde{t}/4)}{\Gamma(1+\alpha'\tilde{t}/4)}\bar{\lambda}_1\Phi_2\Phi_3(k_3\cdot \Gamma)\lambda_4 \, ,
\end{eqnarray}
where we have used the Stirling's formula.

\section{Conclusions}

We have explicitly derived the tree-level two-dilaton two-dilatino scattering amplitude from type II superstring theory by using the Kawai-Lewellen-Tye relations. Then, we have obtained the corresponding Regge behavior.

We would like to emphasize that the relevance of this calculation relies in two important aspects. On the one hand, for the first time the amplitude of two fermions and two bosons for closed superstrings has been explicitly calculated from first principles. This required a significant technical effort. Furthermore, the final expression obtained for the case of two dilatons and two dilatinos is very compact and also explicitly exhibits the ${\tilde{s}}$-${\tilde{u}}$ duality. On the other hand, the results of this calculation allow for applications within the framework of dual gauge/string theory duality, for example, it may be useful to investigate hard scattering of two fermions and two glueballs, alongside the ideas of \cite{Martin:2025jab}. 

As we explained in our previous article \cite{Martin:2025pug}, let us recall that in the GS formalism the vertices associated with massless particle states from open strings have been derived in the light-cone gauge. The emission vertex of a massless fermion whose wave function is given by $u^{\dot{a}}$ and carrying momentum $k^\mu$ is given by \cite{Green:1987sp}:
\begin{equation}
V_F(u, k) = u \ F \ e^{i k \cdot X} = (u^a \ F^a + u^{\dot{a}} \ F^{\dot{a}}) \ e^{i k \cdot X} \, ,
\end{equation}
where
\begin{equation}
F^a =  (p^+/2)^{1/2} S^a \, . \label{Fa}
\end{equation}
Notice that the pre-factor associated with $F^{\dot{a}}$ is $(2 p^+)^{-1/2}$, where $p^+ \propto {\tilde{s}}^{1/2}$. On the other hand, $S^a$'s are dimensionless, which can be seen from the following anti-commutation relation:
\begin{eqnarray}
\{S_m^a, S_n^b\} = \delta_{m+n} \delta^{ab} \, .
\end{eqnarray}
This implies that each fermionic open string emission vertex contains a factor $\tilde{s}^{1/4}$ from $F^a$ as shown in equation (\ref{Fa}). Considering two fermions we have the factor $\tilde{s}^{1/2}$, since each dilatino wave function has a $\tilde{s}^{1/4}$ pre-factor. From the KLT relations, the bosonic kinematic factor carries a factor $\tilde{s}^2$. We can choose $k_3\sim \tilde{s}^{1/2}$. Multiplying all these factors, it leads to $(\alpha'\tilde{s})^{2+\frac{\alpha'\tilde{t}}{2}}$ in the Regge limit, which is understood as the Reggeization of a single graviton in the $\tilde{t}$-channel.

The conclusion is that after taking into account all these elements, scattering amplitudes  $\mathcal{A}_{\text{closed-Regge}}^{\text{4-dilatons}}(k_1,k_2,k_3,k_4)$ and $\mathcal{A}_{\text{closed-Regge}}^{\text{4-dilatinos}}(\tilde{k}_1,\tilde{k}_2,\tilde{k}_3,\tilde{k}_4)$ which we obtained in our previous work  \cite{Martin:2025pug}, and $\mathcal{A}_{\text{closed-Regge}}^{\text{2 dilatons-2 dilatinos}}(\tilde{k}_1,k_2,k_3,\tilde{k}_4)$, which we have obtained in the present work, have the same scaling behavior $(\alpha'\tilde{s})^{2 +\frac{\alpha' \tilde{t}}{2}}$ in the Regge limit. Thus, this trajectory  $J(\tilde{t})=2+\frac{\alpha'\tilde{t}}{2}$ shows the Reggeization of a single graviton in these three cases, as expected since we consider closed strings.

~

~

%
\centerline{\large{\bf Acknowledgments}}
%

\vspace{0.5cm}

The work of L.M., M.P. and M.S. has been supported by the Consejo Nacional de Investigaciones Cient\'{\i}ficas y T\'ecnicas of Argentina (CONICET).

\pagebreak

\newpage

\appendix

%
%
\section{Detailed calculation of the term $\mathcal{T}_{13-1}$}
%
%

As an example we show the explicit calculation of one of the 30 terms obtained from the tensor product of the two open string kinematic factors contributing to the calculation of the scattering amplitude:
\begin{eqnarray}
    \mathcal{T}_{13-1}&=&\frac{\tilde{u}}{2\cdot 16}(\zeta_3\cdot k_4 \zeta_2\cdot k_1 \zeta_1\cdot \zeta_4)\otimes \frac{\tilde{u}}{2\cdot 8}\bar{u}_1\Gamma\cdot \zeta_2 \Gamma\cdot (k_3+k_4)\Gamma\cdot \zeta_3 u_4 \nonumber \\
    &=& \frac{\tilde{u}^2}{512}(\zeta_{3,P} k_4^P \zeta_{2,N} k_1^N \zeta_{1,M} \zeta_4^M)\otimes(\bar{u}_1\Gamma^{N'} \zeta_{2,N'} \Gamma_{M'} (k_3+k_4)^{M'}\Gamma^{P'} \zeta_{3,P'} u_4)\nonumber\\
    &=& \frac{\tilde{u}^2}{512}k_4^Pk_1^N(k_3+k_4)^{M'}\bar{\lambda}_1\Gamma_M\Gamma^{N'}\Gamma_{M'}\Gamma^{P'}\Gamma^M\lambda_4\Theta_{2,NN'}\Theta_{3,PP'}\nonumber\\
     &=& \frac{\tilde{u}^2}{8\cdot 512}k_4^Pk_1^N(k_3+k_4)^{M'}\bar{\lambda}_1\Gamma_M\Gamma^{N'}\Gamma_{M'}\Gamma^{P'}\Gamma^M\lambda_4\Phi_2\Phi_3\nonumber\\
     &\times& (\eta_{NN'}-k_{2,N}\bar{k}_{2,N'}-k_{2,N'}\bar{k}_{2,N})(\eta_{PP'}-k_{3,P}\bar{k}_{3,P'}-k_{3,P'}\bar{k}_{3,P})\nonumber\\
    &=& -\frac{\tilde{u}^2}{8\cdot 512}\Phi_2\Phi_3 (k_3+k_4)^{M'}(k_1-k_2)_{N'}(k_3-k_4)_{P'}\bar{\lambda}_1\Gamma_M\Gamma^{N'}\Gamma_{M'}\Gamma^{P'}\Gamma^M\lambda_4\nonumber\\
    &=&-\frac{\tilde{u}^2}{8\cdot 512}\Phi_2\Phi_3 (k_3+k_4)^{M'}(k_1-k_2)_{N'}(k_3-k_4)_{P'}\bar{\lambda}_1\Gamma_M\Gamma^{N'}\Gamma_{M'}\nonumber\\
    &\times& (-2\eta^{MP'}-\Gamma^M\Gamma^{P'})\lambda_4\nonumber\\
   &=&-\frac{\tilde{u}^2}{8\cdot 512}\Phi_2\Phi_3 [-2\bar{\lambda}_1(k_3-k_4)^{M}\Gamma_M(k_1-k_2)^{N}\Gamma_{N}(k_3\cdot\Gamma)\lambda_4\nonumber\\
   &-&\bar{\lambda}_1\Gamma_M(k_1-k_2)^{N}\Gamma_{N}(k_3+k_4)\cdot \Gamma\Gamma^M(k_3\cdot\Gamma)\lambda_4]\nonumber\\
   &=& -\frac{\tilde{u}^2}{8\cdot 512}\Phi_2\Phi_3[-2(k_3-k_4)^M(k_1-k_2)^N\bar{\lambda}_1(-2\eta_{MN}-\Gamma_N\Gamma_M)(k_3\cdot \Gamma)\lambda_4\nonumber\\
   &-&(k_1-k_2)^N\bar{\lambda}_1(-2\eta_{MN}-\Gamma_N\Gamma_M)(k_3+k_4)\cdot \Gamma\Gamma^M(k_3\cdot \Gamma)\lambda_4]\nonumber\\
   &=& -\frac{\tilde{u}^2}{8\cdot 512}\Phi_2\Phi_3[4(k_3-k_4)\cdot(k_1-k_2)\bar{\lambda}_1(k_3\cdot\Gamma)\lambda_4\nonumber \\
   &+&2\bar{\lambda}_1(k_1-k_2)\cdot\Gamma(k_3-k_4)\cdot\Gamma(k_3\cdot\Gamma)\lambda_4- 2\bar{\lambda}_1(k_1+k_2)\cdot\Gamma(k_1-k_2)\cdot\Gamma(k_3\cdot\Gamma)\lambda_4\nonumber \\
   &+&\bar{\lambda}_1(k_1-k_2)\cdot\Gamma\Gamma_M(k_3+k_4)_N\Gamma^N\Gamma^M(k_3\cdot\Gamma)\lambda_4]\nonumber\\
 &=& -\frac{\tilde{u}^2}{8\cdot 512}\Phi_2\Phi_3[4(\tilde{t}-\tilde{u})\bar{\lambda}_1(k_3\cdot\Gamma)\lambda_4-2\bar{\lambda}_1(k_2\cdot\Gamma)(k_3-k_4)^M\Gamma_Mk_3^N\Gamma_N\lambda_4\nonumber\\
   &-&2\bar{\lambda}_1k_2^M\Gamma_M(k_1-k_2)^N\Gamma_N(k_3\cdot\Gamma)\lambda_4\nonumber \\
   &-&\bar{\lambda}_1(k_2\cdot\Gamma)\Gamma_M(k_3+k_4)_N(-2\eta^{MN}-\Gamma^M\Gamma^N)(k_3\cdot\Gamma)\lambda_4]\nonumber \\
    &=&  -\frac{\tilde{u}^2}{8\cdot 512}\Phi_2\Phi_3[4(\tilde{t}-\tilde{u})\bar{\lambda}_1(k_3\cdot\Gamma)\lambda_4-2(k_3-k_4)^Mk_3^N\bar{\lambda}_1(k_2\cdot\Gamma)(-2\eta_{MN}-\Gamma_N\Gamma_M)\lambda_4\nonumber\\
    &-& 2k_2^M(k_1-k_2)^N\bar{\lambda}_1(-2\eta_{MN}-\Gamma_N\Gamma_M)(k_3\cdot\Gamma)\lambda_4+2\bar{\lambda}_1(k_2\cdot\Gamma)(k_3+k_4)\cdot\Gamma(k_3\cdot \Gamma)\lambda_4\nonumber\\
    &-&10\bar{\lambda}_1(k_2\cdot\Gamma)(k_3+k_4)\cdot\Gamma(k_3\cdot \Gamma)\lambda_4]\nonumber\\
    &=&-\frac{\tilde{u}^2}{8\cdot 512}\Phi_2\Phi_3[4(\tilde{t}-\tilde{u})\bar{\lambda}_1(k_3\cdot\Gamma)\lambda_4+4k_3\cdot(k_3-k_4)\bar{\lambda}_1(k_2\cdot\Gamma)\lambda_4\nonumber\\
    &+&2\bar{\lambda}_1(k_2\cdot\Gamma)(k_3\cdot\Gamma)(k_3-k_4)\cdot\Gamma\lambda_4+4k_2\cdot(k_1-k_2)\bar{\lambda}_1(k_3\cdot\Gamma)\lambda_4\nonumber\\
    &+&2\bar{\lambda}_1(k_1-k_2)\cdot\Gamma(k_2\cdot\Gamma)(k_3\cdot\Gamma)\lambda_4-8\bar{\lambda}_1(k_2\cdot\Gamma)(k_3+k_4)\cdot\Gamma(k_3\cdot\Gamma)\lambda_4]\nonumber\\
     &=&-\frac{\tilde{u}^2}{8\cdot 512}\Phi_2\Phi_3[4(\tilde{t}-\tilde{u})\bar{\lambda}_1(k_3\cdot\Gamma)\lambda_4-4\tilde{s}\bar{\lambda}_1(k_3\cdot\Gamma)\lambda_4-8\bar{\lambda}_1(k_2\cdot\Gamma)k_4^M\Gamma_Mk_3^N\Gamma_N\lambda_4]\nonumber\\
     &=&-\frac{\tilde{u}^2}{8\cdot 512}\Phi_2\Phi_3[4(\tilde{t}-\tilde{u})\bar{\lambda}_1(k_3\cdot\Gamma)\lambda_4-4\tilde{s}\bar{\lambda}_1(k_3\cdot\Gamma)\lambda_4\nonumber\\
     &-&8k_4^Mk_3^N\bar{\lambda}_1(k_2\cdot\Gamma)(-2\eta_{MN}-\Gamma_N\Gamma_M)\lambda_4]\nonumber\\
      &=&-\frac{\tilde{u}^2}{8\cdot 512}\Phi_2\Phi_3[8\tilde{t}\bar{\lambda}_1(k_3\cdot\Gamma)\lambda_4-16k_3\cdot k_4\bar{\lambda}_1(k_3\cdot\Gamma)\lambda_4]\nonumber\\
      &=&-\frac{\tilde{u}^2}{8\cdot 512}\Phi_2\Phi_3[8\tilde{t}\bar{\lambda}_1(k_3\cdot\Gamma)\lambda_4+8\tilde{s}\bar{\lambda}_1(k_3\cdot\Gamma)\lambda_4]\nonumber\\
      &=&\frac{\tilde{u}^3}{512}\bar{\lambda}_1\Phi_2\Phi_3(k_3\cdot \Gamma)\lambda_4.
\end{eqnarray}

\end{document}